%% file: bare_jrnl.tex
\documentclass[journal]{IEEEtran}

\ifCLASSINFOpdf

\else

\fi

\usepackage{wrapfig}
\usepackage{soul}

\usepackage{amsmath,amsfonts}
\usepackage{textcomp}
\usepackage{xcolor}

\usepackage{macros}
\usepackage{subcaption}
\usepackage[font={small,it}]{caption}
\usepackage{color}
\usepackage{tikz}

\usepackage{subcaption}
\usepackage{graphicx}
\usepackage{tabularx}
\usepackage{array,multirow,rotating}
\usepackage{algorithm}
\usepackage{algpseudocode}
\usepackage{tikz}
\usepackage{hyperref}
\usepackage{adjustbox}
\usepackage{float}

\usepackage{mathtools}
\usepackage{lipsum}

\usepackage{macros}

\usepackage{pifont}

\usepackage[compact]{titlesec}

\usepackage[inline]{enumitem}

\def\footnoterule{\kern-3pt
  \hrule \kern 2.6pt} 
            
\usepackage{wrapfig}

\def\BibTeX{{\rm B\kern-.05em{\sc i\kern-.025em b}\kern-.08em
    T\kern-.1667em\lower.7ex\hbox{E}\kern-.125emX}}

\AtBeginDocument{%
  \providecommand\BibTeX{{%
    \normalfont B\kern-0.5em{\scshape i\kern-0.25em b}\kern-0.8em\TeX}}}

\usepackage{setspace}

\begin{document}

\title{Seeker: Synergizing Mobile and Energy Harvesting Wearable Sensors for Human Activity Recognition}

\author{Cyan Subhra Mishra,~\IEEEmembership{Student Member,~IEEE,}
        Jack Sampson,~\IEEEmembership{Member,~IEEE}
        ~Mahmut Taylan Kandemir,~\IEEEmembership{Fellow,~IEEE}
        and~Vijaykrishnan Narayanan,~\IEEEmembership{Fellow,~IEEE}\\
        The Pennsylvania State University, USA. (Email: \{cyan, jms1257, mtk2, vijaykrishnan.narayanan\}@psu.edu) \vspace{-16pt}}

\maketitle
\setstretch{0.9775}
\begin{abstract}
\label{sec:abstract}
\input{src/00abstract}
\end{abstract}


\IEEEpeerreviewmaketitle

\section{Introduction}
\label{sec:introduction}
\input{src/01Introduction}

\section{Background and Motivation}
\label{sec:bg-rw}
\input{src/02BG-RW}

\section{Design Space Exploration}
\label{sec:motivation}

\input{src/03DesignSpaceExp}

\section{\emph{Seeker} - A Synergistic Sensor-Host Ecosystem}
\label{sec:4}
\input{src/04Seeker}

\section{Evaluation and Results}
\label{sec:evaluation}
\input{src/05Evaluation}

\section{Conclusion}
\label{sec:conclusion}
\input{src/06Conclusion}

\ifCLASSOPTIONcaptionsoff
  \newpage
\fi

\bibliographystyle{IEEEtran}
\bibliography{References}

\end{document}

%% file: src/00abstract.tex
There is an increasing demand for intelligent processing on emerging ultra-low-power internet of things (IoT) devices, and recent works have shown substantial efficiency boosts by executing inference tasks directly on the IoT device (node) rather than merely transmitting sensor data. However, the computation and power demands of Deep Neural Network (DNN)-based inference pose significant challenges for nodes in an energy-harvesting wireless sensor network (EH-WSN). Moreover, these tasks often require responses from multiple physically distributed EH sensor nodes, which imposes crucial system optimization challenges in addition to per-node constraints.  

To address these challenges, we propose \emph{Seeker}, a novel approach to efficiently execute DNN inferences for Human Activity Recognition (HAR) tasks, using both an EH-WSN and a host mobile device. Seeker minimizes communication overheads and maximizes computation at each sensor without violating the quality of service. \emph{Seeker} uses a \emph{store-and-execute} approach to complete a subset of inferences on the EH sensor node, reducing communication with the mobile host. Further, for those inferences unfinished because of harvested energy constraints, it leverages an \emph{activity aware coreset} (AAC) construction to efficiently communicate compact features to the host device where ensemble techniques are used to efficiently finish the inferences. \emph{Seeker} performs HAR with $86.8\%$ accuracy, surpassing the $81.2\%$ accuracy of a state of the art approach. Moreover, by using AAC, it lowers the communication data volume by $8.9\times$.  

%% file: src/01Introduction.tex
Innovations in low-power computing, artificial intelligence, and communication technologies have given rise to the generation of intelligent connected devices which constitute the Internet of Things (IoT). While IoT platforms are diverse, mobility is common and wearables are an increasingly important class of platforms:
For instance, the number of connected wearable devices is expected to increase to 1,105 million in 2022~\cite{Cisco-report}. At the same time, application demands for these devices are also growing: IoT devices, including wearables, are now expected to participate in producing rapid inferences as part of the increasingly complex tasks enabled by machine learning algorithms~\cite{netadapt}.
Furthermore, considering the role of wearable devices, especially in wireless body area networks, emerging applications and research are now targeted towards optimizing the wearable devices, making them more user friendly and more intelligent. These devices are now expected to be capable of handling many complex tasks, such as human activity recognition (HAR), machine translation, biometric authentication, fall and accident detection. 

However, given their form-factor constraints, it is challenging to execute compute-intensive inference tasks directly on wearable platforms due to their limited energy storage and low-power operation.
Therefore, complex inference tasks, such as human activity recognition (HAR)  
are typically offloaded either to the cloud or a connected (to internet) host device, such as a smartphone. 
To alleviate energy provisioning challenges, recent works~\cite{IntBeyondEdge, chinchilla} proposed energy harvesting, along with multiple compiler and runtime optimizations, to increase local compute at the edge. Furthermore, non-volatile processors (NVP)~\cite{NVPMa} have been proposed to maximize the forward progress when utilizing harvested energy sources.

Prior work~\cite{Origin} has explored methods for combining intelligent scheduling and ensemble learning to shift inference computations onto EH sensor nodes for distributed wearable tasks, such as HAR. However, given the limitations of an EH budget, such approaches usually end up dropping many samples entirely, neither inferring from them locally nor transmitting the raw data due to a lack of sufficient energy. Conventional edge devices have relied on compression techniques to minimize data communication overheads and help cover this gap. However, when applied to low dimensional sensor data,  classical lossy compression techniques tend to discard important features to achieve a higher compression ratio, which significantly degrades the inference accuracy. To mitigate this problem, recent works~\cite{bachem2015coresets, Ting-He-ArXiv, Ting-He-IEEE} propose using \textit{coresets}, a data representation technique from computational geometry that preserves important, representative features when building a compressed form of the data, for efficient edge communication. However, performing accurate inference on coresets can still be challenging.

This paper proposes \emph{Seeker}, a novel approach that leverages and extends coresets to efficiently execute DNN inference, specifically HAR, across a set of EH-WSN nodes and a host mobile device. \textit{Seeker} builds upon previously proposed~\cite{Origin} HAR ensembling for EH-WSNs, and applies coreset techniques to both increase the number of inferences that can be performed locally and make transmission viable in more cases where computation cannot be performed on-node. Moreover, Seeker augments its coreset formation with application-awareness and provides hardware support for coreset formation acceleration to make them sufficiently computationally efficient, adaptive, and accuracy-preserving for low-power, latency critical embedded applications.
In summary, this paper makes the following \textbf{primary contributions}:

\begin{enumerate}[leftmargin=*]

\item We develop extensions to traditional coreset formation that enhance their applicability to EH-WSN inference scenarios. Specifically, we introduce an \emph{activity-aware coreset construction} technique to dynamically adapt to both the activity and the available harvested power, while conserving maximum features of the data. We also propose a \emph{recoverable coreset construction} technique, which helps reconstruct the original data from the compressed form with minimum \textit{inference loss}. This combination achieves superior accuracy-compression tradeoffs compared to classical compression techniques like DCT and DWT.

\item We propose simple, low power, and low latency hardware to efficiently build coresets from the raw data, further increasing the number of samples that can be inferred or transmitted under a fixed EH budget. 

\item We design augmentations for the sensors to employ data memoization for skipping unnecessary compute, and the use of low precision ReRAM x-bar based DNN-engines for energy efficient inference. 

\item We provide a detailed evaluation of our system and the proposed hardware design. Our evaluation shows that, even when powered by an unreliable EH source, {\em Seeker}'s coreset-based optimizations result in better accuracy than that of a fully-powered system running a state-of-the-art classifier optimized for energy efficiency. Specifically, \emph{Seeker} reaches  86.8\% top-1 accuracy in comparison to the 81.2\% accuracy of the baseline system, while reducing the data communication by $\approx 8.9\times$ compared to the state-of-the-art optimization techniques proposed for EH-WSNs.   
\end{enumerate}

\begin{figure*}
\centering
    \hfill
	\subfloat[HAR in a conventional setting]
	{
	\includegraphics[width=0.4\linewidth, height = 6cm]{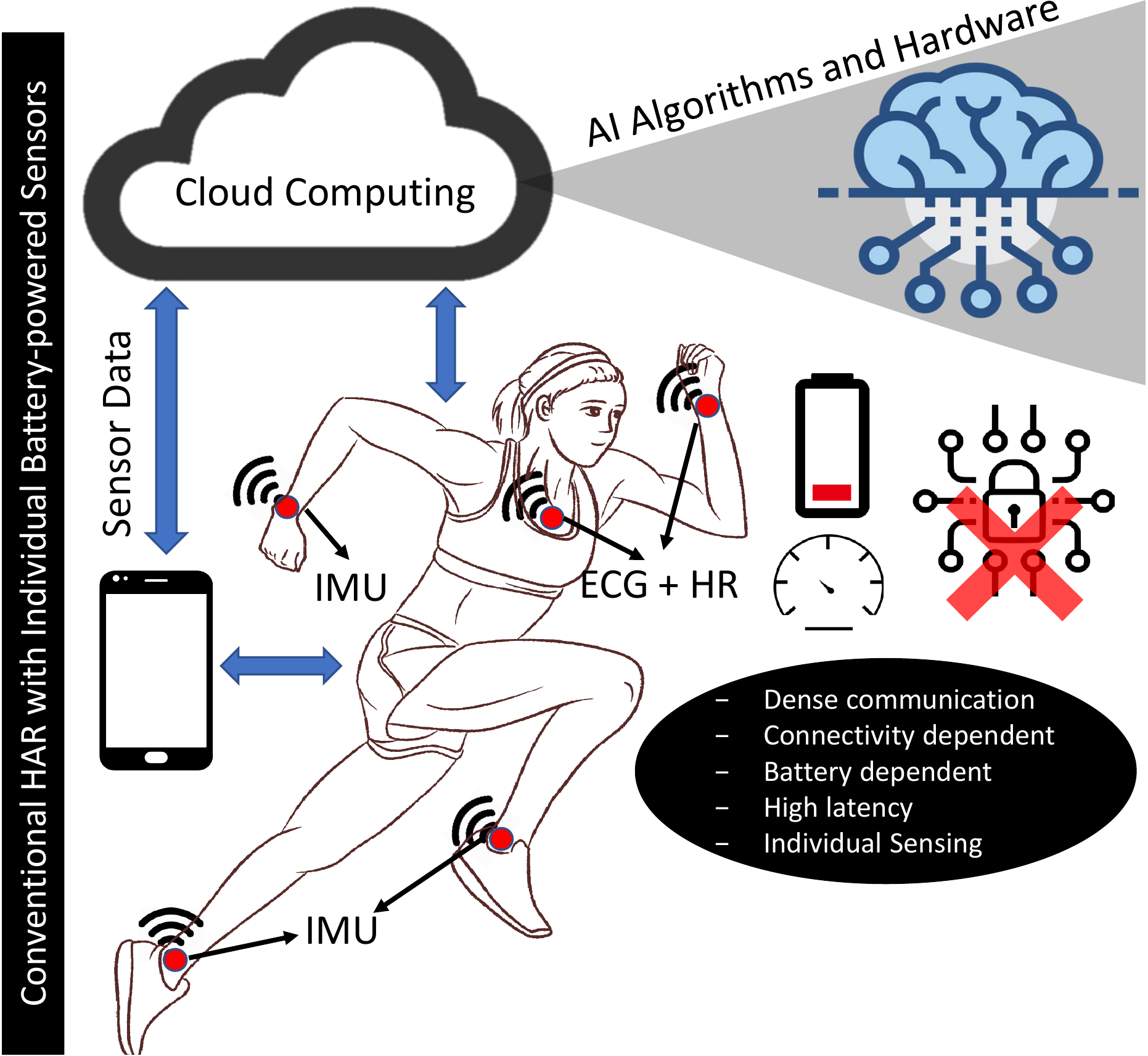}\label{Fig:introFig-a}
	}
	\hfill
    \subfloat[On-device HAR using EH-WSNs]
    {
    \includegraphics[width=0.4\linewidth, height = 6cm]{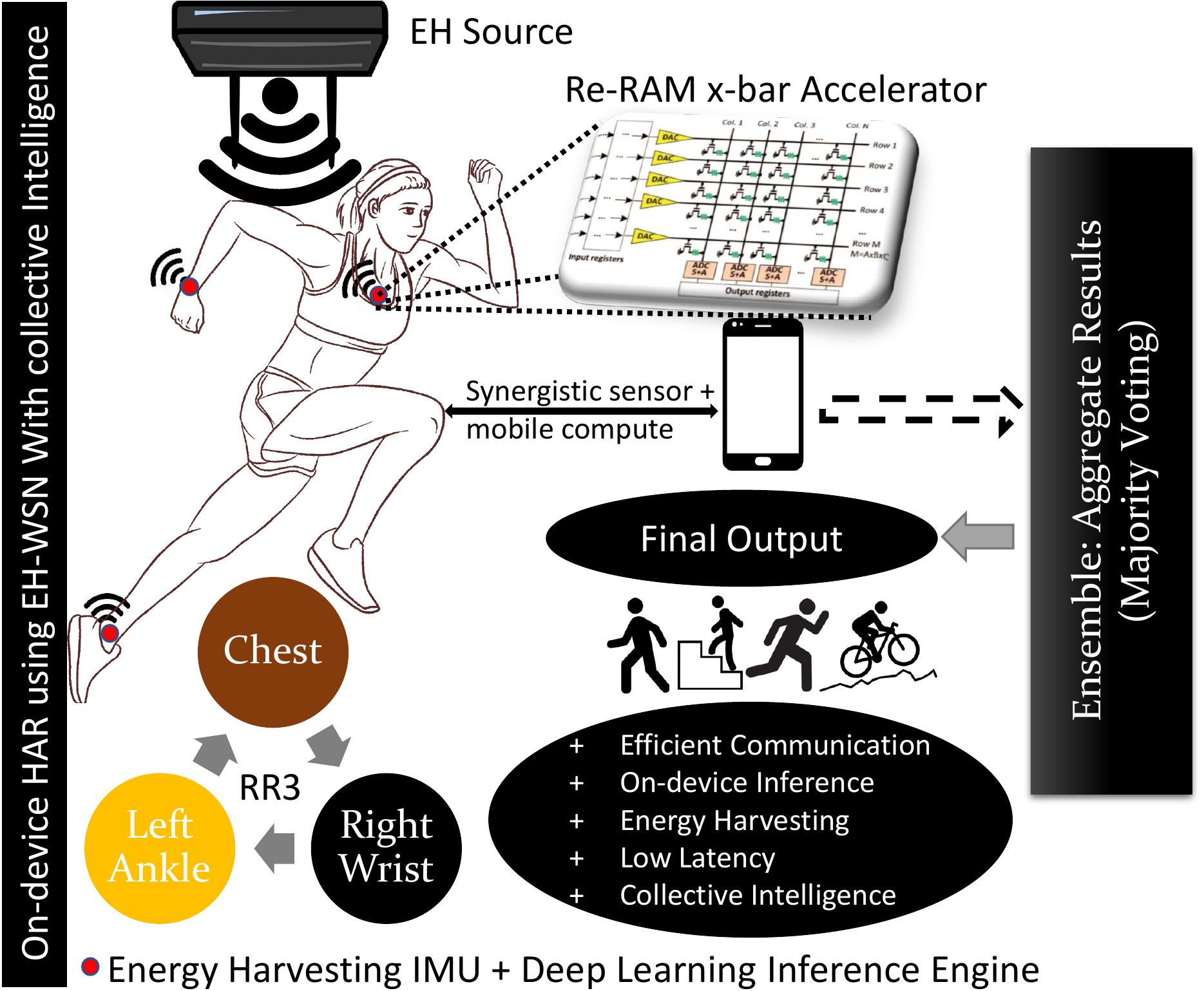}\label{Fig:introFig-b}
    }
    \hfill
    \vspace{-0.1cm}
    \caption{A representation of edge computation (with Human Activity Recognition (HAR) as an example). Fig.~\ref{Fig:introFig-a} represents the conventional setup where the barely intelligent battery backed edge devices/ sensors stream data to the host or cloud for inference which is inefficient in terms of both latency and power. Fig.~\ref{Fig:introFig-b} shows a desired set-up, where the energy harvesting wireless sensor network work in cohesion to get a more accurate inference while maximizing compute at the edge and making minimum and efficient communication with a host.}\vspace{-6pt}
    \label{Fig:introFig}  
\end{figure*}

%% file: src/02BG-RW.tex
In this section, we describe how current low-power sensor networks, especially wireless body area networks, work and how they perform complex inference tasks like HAR. We specifically detail the role played by energy harvesting techniques and Non Volatile Processors (NVP) in HAR. 
\subsection{Human Activity Recognition at the Edge}

\label{subsec:BG-EHWS}
Sensors and sensor networks comprise much of the IoT and continue to dominate its growth. They have become an integral part of daily life in the form of smart watches, smart glasses, and even smart home appliances. A large portion of these devices are mobile and battery operated, especially wearable and healthcare devices~\cite{Cisco-report}. Furthermore, these devices are performing increasingly complex computations and engaging in frequent communication with each other, with a master device, or with the cloud for synchronization, compute and updates. However, the small form factor of these devices leads to limited battery life, which constrains the quantity of computation performed between charges~\cite{NVPMa}. Furthermore, scaling the number of such battery operated devices is challenging from economical, logistical and environmental prospective~\cite{incidental}. Therefore, many recent works~\cite{NVPMa, incidental, chinchilla, LuciaCheckpoint} propose using energy harvesting as a solution for operating these devices. The fickle nature of harvested energy has, in turn, lead to a focus on check-pointing, code optimization and compiler level tweaks help maximizing the forward progress on such devices~\cite{incidental, chinchilla, LuciaCheckpoint}. From a hardware point of view, these works rely on a small capacitor as an intermediate storage for the harvested energy and mostly used off-the-shelf commercial hardware platforms, while relying on software optimizations and judicious use of persistent storage to complete smaller workloads like \textit{"Ok Google Detection"}. 

\label{subsec:BG-NVP}
Although software solutions can be sufficient for some deployments, they are not efficient for all types of programs and for all types of harvested energy sources: Specifically, for very weak sources, it becomes challenging to perform frequent check-pointing in software while making any progress on computation. 
Therefore, to reduce the effort, energy and time spent in the save-and-restore paradigm proposed in~\ref{subsec:BG-EHWS}, recent works~\cite{NVPMa, incidental, NVPMicro, spendthrift, ResiRCA} propose use of a non-volatile processor, where the non-volatility of the hardware itself takes care of saving and resuming the program execution. This reduces software overheads and latencies for handling power emergencies and hence can guarantee better QoS for complex and longer tasks even when power is deeply unreliable. Using an NVP, and multiple harvested energy sources~\cite{ResiRCA} demonstrates the possibility of performing complex DNN inference at the EH-Sensor itself.

While an NVP ensures safe check-pointing for a given computation, current edge scenarios may require a device to be simultaneously performing multiple functionalities~\cite{iWatchBattery, AppleFall, AppleECG, Google-Assistant-Watch}. As a result, it is difficult to reliably run all these complex task standalone on the edge device. Current devices adapt in one of three ways: \emph{1) Send all the sensor data to a connected host device, or cloud, to offload the compute and act only as a sensing and display device (Fig.~\ref{Fig:introFig-a}). 2) Process data on the device itself, potentially dropping or delaying tasks due to energy shortfalls (\ref{Fig:introFig-b} ).  3) A mix of the two models, where some computations do happen on the device while others are offloaded to balance compute, energy, and communication resources.} Typically, the latter is preferred, but it is non-trivial to find the optimal balance between what is to be done on the edge and what should be offloaded~\cite{kang2017neurosurgeon, taylor2018adaptive, zhao2018deepthings, eshratifar2018energy}. Since recent works~\cite{spendthrift,ResiRCA} show processing at the edge is more efficient, when there are appropriate acceleration resources, we will explore strategies that push more compute to the edge sensors and rely less on more powerful coordinating devices.

\subsection{Motivation}

\begin{figure*}[h]
\centering
	\subfloat[Completion with Extended-Round-Robin]
	{
	\includegraphics[width=0.30\linewidth, height=2.5cm]{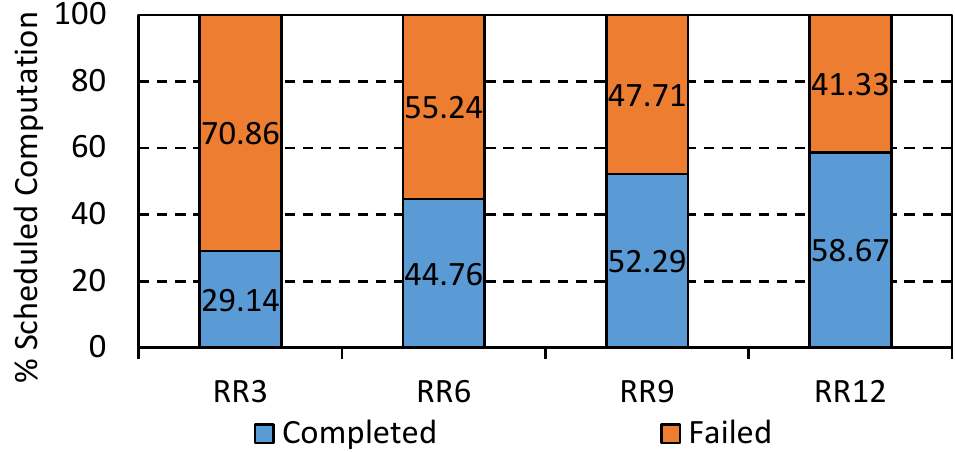}\label{Fig:frac-complete}
	}
	\hfill
    \subfloat[Accuracy of Extended-Round-Robin]
    {
    \includegraphics[width=0.30\linewidth, height=2.5cm]{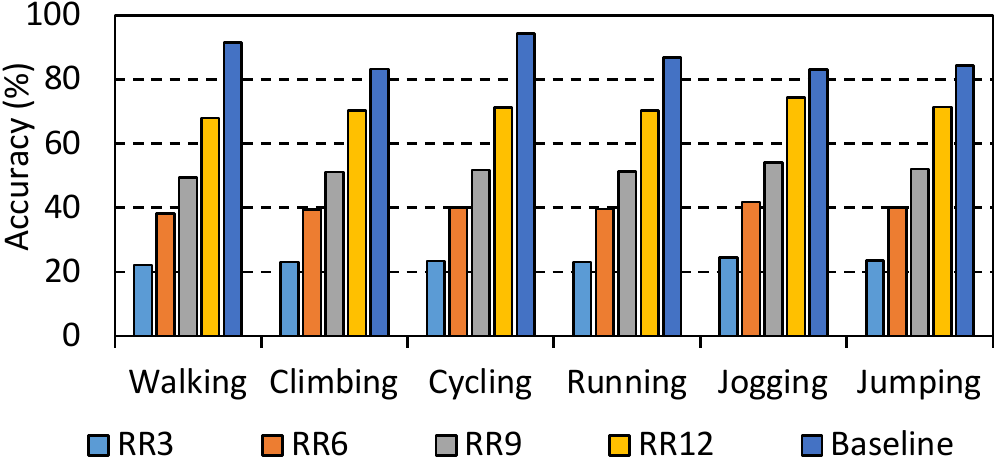}\label{Fig:RR-accuracy}
    }
    \hfill
    \subfloat[Accuracy with different quantizations]
    {
    \includegraphics[width=0.30\linewidth, height=2.5cm]{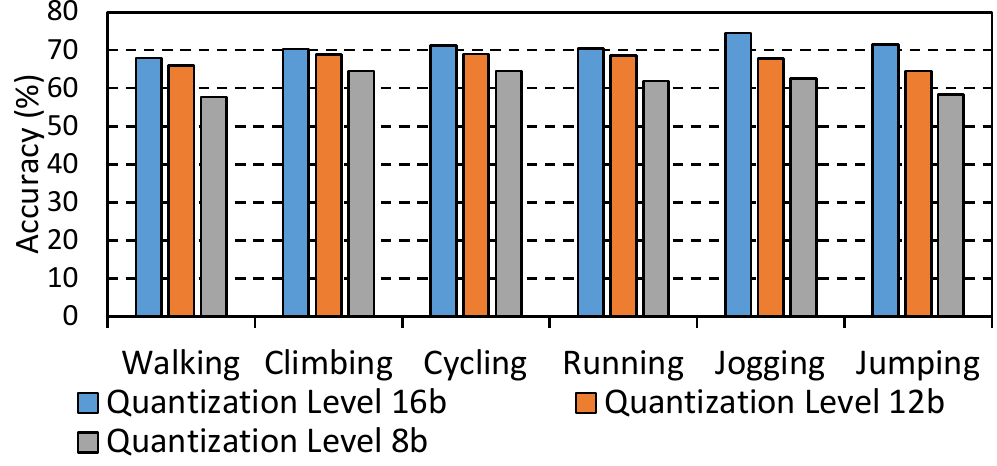}\label{Fig:quantized-accuracy}
    }
    \caption{Accuracy comparison of various classical node-level optimization techniques. The Extended-Round-Robin policy~\cite{Origin} takes a store-and-execute approach where computations are performed in the sensors in a round-robin fashion, and the number associated represents the ratio of store cycles vs execute cycles (e.g. RR3 is 3 store cycles followed by 1 execute cycle). The 'Baseline' model is a fully powered system with no energy restrictions, and the quantized model runs on harvested energy using a store and execute fashion with a RR12 policy.}
    \label{fig:Projection}\vspace{-8pt}  
\end{figure*}

\begin{table}[h]
\vspace{-8pt}\centering
\begin{adjustbox}{max width=\linewidth}
\begin{tabular}{|c|c|c|}
\hline
\textbf{Algorithm}    & \textbf{Compression Ratio} & \textbf{Accuracy Loss (\%)} \\ \hline
Fourier Decomposition & 3 - 5                      & 9.1 - 18.3                  \\ \hline
DCT                   & 3 - 5                      & 5.8 - 16.2                  \\ \hline
DWT                   & 3 - 6                      & 5.3 - 12.7                 \\ \hline
Coreset               & 3 - 10                     & 0.02 - 0.76                 \\ \hline
\end{tabular}
\end{adjustbox}
\caption{Accuracy trade-off of different compression techniques: Low-dimensional data loses important features under lossy compression, dropping inference accuracy significantly compared to the original data. Details on Coreset are available on Section~\ref{sec:4}.}
\label{tab:compression-table}
\end{table}

The compute and power constraints of the EH systems, and the real-time nature of HAR, makes DNN inference quite challenging on EH-WSNs. 
Since communicating raw sensor data to a host device is expensive, several proposed  systems~\cite{Origin,intelligenceBeyondEdge, ResiRCA} prefer to execute the inferences at the edge sensors. However, completing all inferences with a harvested power budget is challenging. To quantify the scope of this challenge, we perform experiments on the MHEALTH data-set~\cite{mHealth,mHealthDroid} (see Section~\ref{sec:evaluation} for data-set details) using the DNNs proposed in~\cite{HARHaChoi,HARHompel}, an energy harvesting friendly DNN hardware accelerator~\cite{ResiRCA} and recently proposed HAR-specific optimizations for EH systems~\cite{Origin}. Fig.~\ref{Fig:frac-complete} shows that a state-of-the-art system still only finishes $\approx 58.7\%$ of the inferences scheduled on a sensor. Although Fig.~\ref{Fig:RR-accuracy} shows how accuracy can increase by further tuning duty-cycle, the returns are diminishing. This system does not aggressively employ quantization, which is a commonly used technique~\cite{drq-isca2020} to reduce both compute and transmission energy in DNN tasks, and Fig.~\ref{Fig:quantized-accuracy} shows accuracy as a function of quantization (we took the approach of performing post training quantization and fine-tuned the DNN to work with reduced bit precision instead of training the DNN from scratch with a reduced precision). The quantized DNNs benefit from lower compute and memory footprints, but need specialized fine-tuning and often suffer from lower accuracy, as do other  approximation-via-data-reduction techniques such as sub-sampling. Collectively, the aforementioned figures demonstrate that the harvested energy budget is insufficient to perform {\em all} inferences with acceptable accuracy on currently proposed EH-WSN systems.

\textbf{Data Compression:~}
Since performing the inference on the sensor node and communicating the result takes \textit{\textbf{less}} energy (for the DNNs in question) than communicating the raw data~\cite{NVPMa, ResiRCA, incidental}, communicating the \textbf{raw} data in cases where there is insufficient energy to perform inference is not a viable option. 
One approach to combat the high cost of transmission, common in EH-WSNs, is compression. Using standard compression algorithms, like discrete cosine transform and Fourier decomposition, to minimize the communication overhead is also not a viable solution~\cite{compression-marcelloni2008simple}: This is partly because we need a very high compression ratio with very low power, and secondly, these compression algorithms are not context-aware, resulting in degraded inference accuracy (see TABLE~\ref{tab:compression-table} for details). A key insight is that, while these compression techniques work very well for high dimensional data (e.g. images), inference on low-dimensional sensor data is much more sensitive to lossy compression and will not achieve sufficient compression ratios from lossless approaches.  
Thus, our deployment scenario demands a smaller representative form of the data that still \textit{preserves enough  application-specific features to perform meaningful classifications in a given DNN}. 

\textbf{Why Coresets?~}
The aforementioned requirements motivate us to consider \emph{coresets}  for forming representations of the original sensor data. Coresets were primarily used in computational geometry~\cite{bachem2015coresets}, but there have been recent efforts~\cite{Ting-He-ArXiv, Ting-He-IEEE} to use coresets for machine learning and sensor networks. Since coresets can be crafted to preserve features required by subsequent inference stages, they are an effective way to construct a representation of the data set with high compression ratios without incurring unacceptable accuracy losses. For the DNNs in question, coresets can achieve sufficient compression ratios to make communication energy-competitive with computation, as well as opening up new opportunities for optimizing DNN inference on the coreset, rather than original data.  

\textbf{Challenges with Coresets for DNNs:~} Although there have been prior efforts to optimize and quantize coreset construction~\cite{Ting-He-IEEE} for minimizing communication overheads, there remain open questions in using coresets for wearable DNN inference systems: 
1) Can we vary the compression ratio (by forming dynamic coresets) depending upon the complexity of the input data, thereby avoiding unnecessary communication?
2) For simpler coreset construction techniques like clustering, is there any possibilities of reconstructing the original data from the coresets, which is useful for performing inference on a low dimensional sensor data?
3) Given the simpler nature of these algorithms, can we design simple and inexpensive hardware to accelerate the coreset formation while requiring little energy and area to do so?



%% file: src/03DesignSpaceExp.tex
As discussed earlier, the major challenge faced while performing HAR on EH-WSNs is the inability to complete all necessary computations at the node itself. As a result, we need the support of a host device to synergistically distribute computation between the host and the EH sensor-node depending on the harvested energy budget. Since data communication likewise consumes substantial power, we we rely on constructing coresets as an efficient way to lossily communicate the features with minimal information degradation. However, the coreset construction techniques need to be extremely lightweight while preserving key features to justify the computation-communication tradeoffs in energy and latency. To this end, we explore two different kinds of coreset construction techniques that fit these requirements. 
\subsection{Coreset Construction Techniques}
\begin{figure}

  \centering 
  \includegraphics[width=0.95\linewidth]{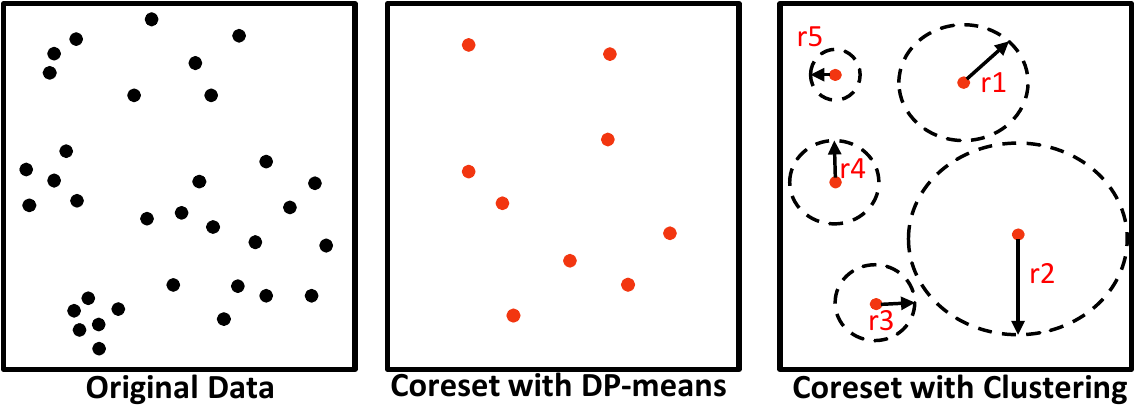}
  \caption{Coreset construction techniques adapted in \emph{Seeker}. DP-means uses a probability based sampling method, whereas clustering tries to preserve the geometric shape of the original data. The points/values in red are communicated to the host.}
  \label{Fig:Coresets}\vspace{-16pt}
\end{figure}
\textbf{Coreset Construction Using DP-means:~} One of the easiest way to build a representative form from a data 
distribution is to perform importance sampling, i.e. choose the data which are unique. 
Coreset construction using  DP-means~\cite{bachem2015coresets} uses Dirichlet process sample data by giving more importance to data-points which are farther enough from each other to build a representative form.
DP-means based coreset construction is computationally inexpensive as it just uses the distance between points to build a probability distribution and then selects the most probable data points. It is therefore very useful for energy-scarce situations where the EH budget is not enough to perform inference and just enough energy is available to communicate some minimal data to the host. The host can take the sub-sampled data and perform inference. Although the inference accuracy may degrade because of sub-sampling, it still outperforms classical compression techniques. 

\textbf{Coreset Construction Using Clustering:~} Although DP-means based coreset construction is computationally inexpensive, it suffers from accuracy loss because it doesn't explicitly preserve the organization of the data points. To address this, we also utilize coreset construction using k-means clustering, which separates the data points into a set of k (or fewer) N-spherical clusters and represents the geometric shape of the data by using the cluster centers and cluster radii (Fig.~\ref{Fig:Coresets}). These are then communicated to the host device for inference. We observe that inferences with coresets constructed using clustering are more accurate than using DP-means, and therefore prefer the former whenever there is enough energy. 

\subsection{Balancing Coreset Communication and Accuracy}

Coreset construction techniques allow a variable number of features depending on the available energy. This reduces the chance that neither data nor an inference will be sent. Although coresets preserve important features, they are lossy, and inference accuracy drops when performed on coresets. 
\begin{wrapfigure}{l}{0.50\linewidth}
  \includegraphics[width=\linewidth]{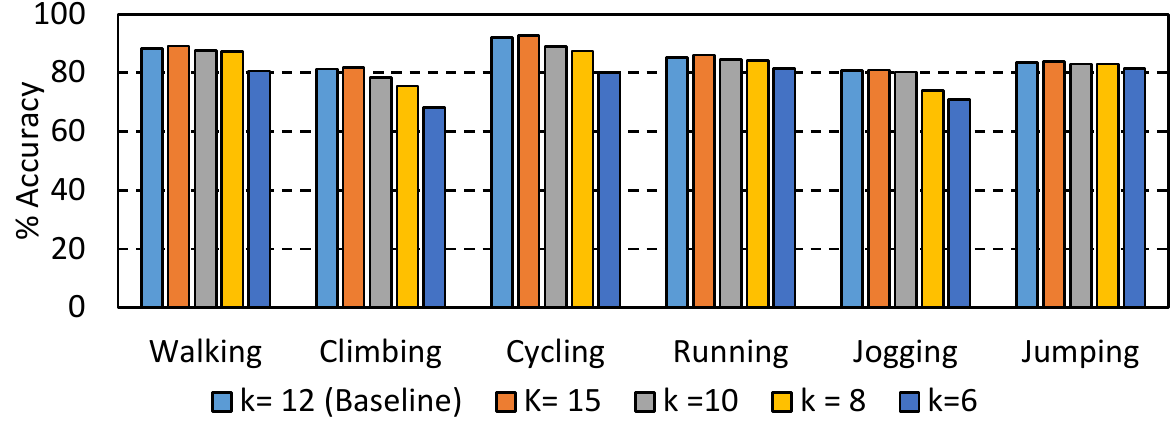}
  \caption{Accuracy of inference when coreset is formed with different number of clusters (k represents the number of coreset clusters).}
  \label{Fig:AAC-Accuracy}
  \vspace{-3mm}
\end{wrapfigure}
This leaves an optimization space in trading between communication cost vs. accuracy, i.e. whether to construct strict and low-volume coresets and lose accuracy or to preserve maximum data points and pay for the communication cost. We perform an analysis on the MHELATH~\cite{mHealth, mHealthDroid} data set (we take a overlapping moving window of 60 data points sampled at 50Hz from 3 different IMUs, overlap size: 30 data points) to find a tradeoff between the coreset size (directly related to the communication cost) and the inference accuracy. Empirically, we observe that accurately preserving the features for each activity requires \textbf{20 data points} using DP-means or \textbf{12 clusters} (see Fig.~\ref{Fig:AAC-Accuracy}) using clustering based techniques. Going above 12 clusters did not significantly improve accuracy, while going below 12 caused significant accuracy drops.

Since the DNN models were originally designed to infer on the full data, we retrain the DNN models to recognize the compressed representation of the data and classify the activity directly from that (both from the DP-means and clustering). As the coreset formation algorithms are fairly simple~\cite{bachem2015coresets, practicalcoresets, Ting-He-ArXiv, Ting-He-IEEE}, it does not take much latency or energy to convert the raw sensor data into the coreset form. This allows the sensor to opt for coreset formation followed by data communication to the host device as an energy-viable alternative to local DNN inference on the original data. While transmitting the raw data (60 data points, 32bit-fp data type) needs 240Bytes of data transfer, with coreset construction and quantization we can limit it to 36Bytes (for 12 clusters, each cluster center is represented by 2Bytes of data, and radius represented by 1Byte data), and thereby reducing the data communication volume by 85\%. The host runs inference on the compressed data to detect the activity (with an accuracy of 76\%). However, due to this reduced accuracy, the sensor only takes this option when it does not have enough energy to perform the inference at the edge device (either in the 16bit or 12bit variant of the DNN).

\subsubsection{Recoverable Coreset Construction:} The primary reason the accuracy of inferring on coreset data is lower than that of the original model is the loss of features. Since HAR sensor data are low dimensional, even with a good quality of coreset construction, it is difficult to preserve all the features. However, while inferring at the host, if we are able to recover the data or reconstruct it with minimum error, the accuracy can easily be increased. Since DP-means performs weighted sub-sampling, we cannot recover the unsampled points, but clustering preserves the geometry of the shape and hence we can approximate the missing data points to fit the shape. However, to achieve this, we need some extra information about the clusters. The standard method of clustering-based coreset construction keeps the cluster center and cluster radius, which gives the geometrical shape of the entire data. Extending this with the point count for each cluster allows for reconstruction of data in the original form that can be processed by DNNs trained on full-size data. These reconstructed data sets can be synthesized simply by uniformly distributing the points within each cluster. Although the intra-cluster data distribution will be different from the original, it will still preserve the overall geometry.

From our experiments, we observe that inferring on the synthesized reconstructions of cluster based coresets can achieve an accuracy of $\approx 85\%$. The addition of the reconstruction feature at the host comes with little overhead. The addition of the recovery parameter (number of points per cluster) demands 4 more bits (in our experiments, we never observe any clusters having more than 16 data points) of data per cluster, bringing the total data communication volume to 42Bytes, which is still a significant reduction in comparison to the 240Bytes needed to communicate the raw data. However, since clustering based coreset construction is more expensive than the DP-means based coreset construction, it is not always possible to build a recoverable coreset at the edge.  

\subsubsection{Activity Aware Coreset Construction:}
\label{subsubsec:AAC}
Although cluster-based coreset formation algorithms do an excellent job  of preserving key features, the number of clusters required for computing the coreset depends upon the complexity of the data distribution, that is, for HAR, the complexity of the activity. \begin{wrapfigure}{l}{0.30\textwidth}
  \vspace{-8pt}\includegraphics[width=0.30\textwidth]{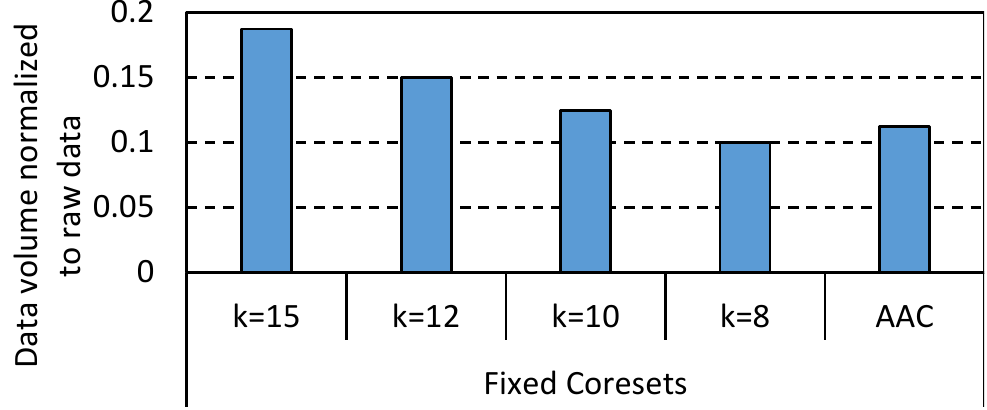}
  \caption{Communication data-volume normalized to full raw data. We show the communication data volume at various cluster sizes and with AAC}
  \label{Fig:AAC-Communication}\vspace{-12pt}
 \end{wrapfigure}In our experiments, we observe that not all the activities are equally complex and hence may or may not need a certain number of clusters to represent every feature.
Fig.~\ref{Fig:AAC-Accuracy} shows the accuracy of each activity when represented with a different number of clusters.  While activities like walking and running do not lose much accuracy even when represented with eight clusters, complex activities are more sensitive. As the communication overhead depends on the number of clusters, which, in turn, depends on the complexity of the activity, we propose an \emph{activity aware clustering} which ensures that coresets for the current activity are represented with a sufficient number of clusters to preserve accuracy, but no more than required. 

 We determine the number of clusters required as a function of current energy availability and accuracy trade off of using a lesser number of clusters. However, naively framed, this approach requires knowledge of what activity is being performed in order to encode the data that will be used to perform inference to determine what activity is being performed. To break this circular dependency, we take inspiration from prior work in HAR~\cite{Origin}, and use the highly stable temporal continuity of human activity (relative to the tens of ms timescales for HAR inferences) to predict the current activity based on previously completed local inferences. We use temporal continuity to our advantage, and make sure that if the system does not have enough energy to form the default 12 clusters, it will resort to forming a smaller number of clusters with minimum accuracy loss. We implement a small lookup table to carry the information on accuracy trade-off for different activities with respect to the number of clusters used to form the coresets (similar to the data represented Fig.~\ref{Fig:AAC-Accuracy}). Fig.~\ref{Fig:AAC-Communication} shows the communication data volume (which is also proportional to communication energy and latency) normalized to sending the raw data using IEEE 802.15.6 protocol. We can see that AAC communicates about \textbf{11\%} data compared to sending the full raw data. Note that we only resort to activity awareness while forming coresets using clustering, as DP-means coreset construction does not require much energy. Furthermore, dropping the number of samples in DP-means method significantly hampers the accuracy, which is not the case for recoverable-clustering based coreset construction.

\subsubsection{Data Memoization:}
\begin{wrapfigure}{r}{0.60\linewidth}
  \vspace{-16pt}\includegraphics[width=\linewidth]{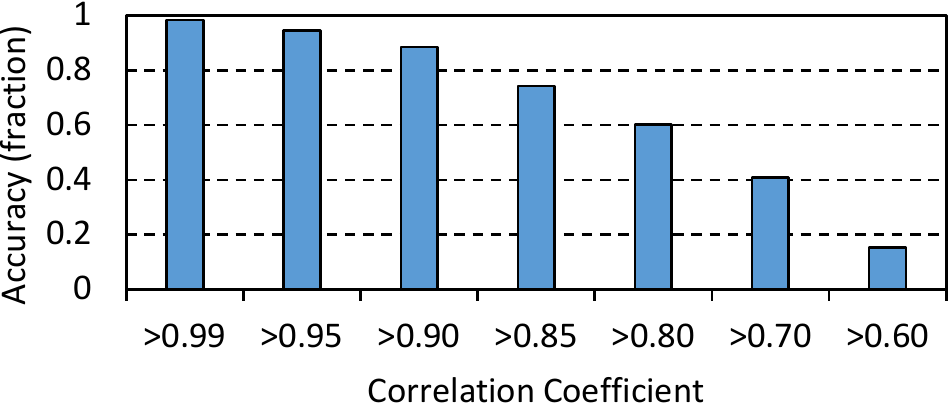}
  \vspace{-16pt}\caption{The probability of two activities being same with respect to the correlation between the sensor data}
  \label{Fig:correlation}
 \end{wrapfigure}
 Given our focus on ultra low power energy harvesting devices, any opportunities to reduce computation and communication can noticeably augment the performance and efficiency of the entire system. We further leverage the temporal continuity of the human activity to skip repetitive inferences.  From \ref{subsubsec:AAC} we know that there is a good probability that the the current activity being performed is same as the activity performed, and hence there is a good probability that the sensor data pattern observed in both these cases bear similar signature. In fact, for two instances of the same activity, there should be a correlation in the sensor data. We empirically measure this by testing for correlation between the sensor signatures of different activities.  
 
 Fig.~\ref{Fig:correlation} shows the probability of two activities being the same with respect to the observed correlation coefficient. Conservatively, we choose a correlation coefficient $\geq 0.95$ to predict that the two activities are the same, and hence skip the inference altogether and just communicate only the results to the host. We store a ground truth sensor data pattern for all the activities, and when a new data arrives, we find the correlation of the sampled data against the ground truth data, and if any of the correlation coefficient comes out to be $\geq 0.95$, we choose to ignore further inference computation and only communicate the classification result to the host for further processing.

%% file: src/04Seeker.tex
\begin{wrapfigure}{l}{0.60\linewidth}
  \includegraphics[width=0.9\linewidth]{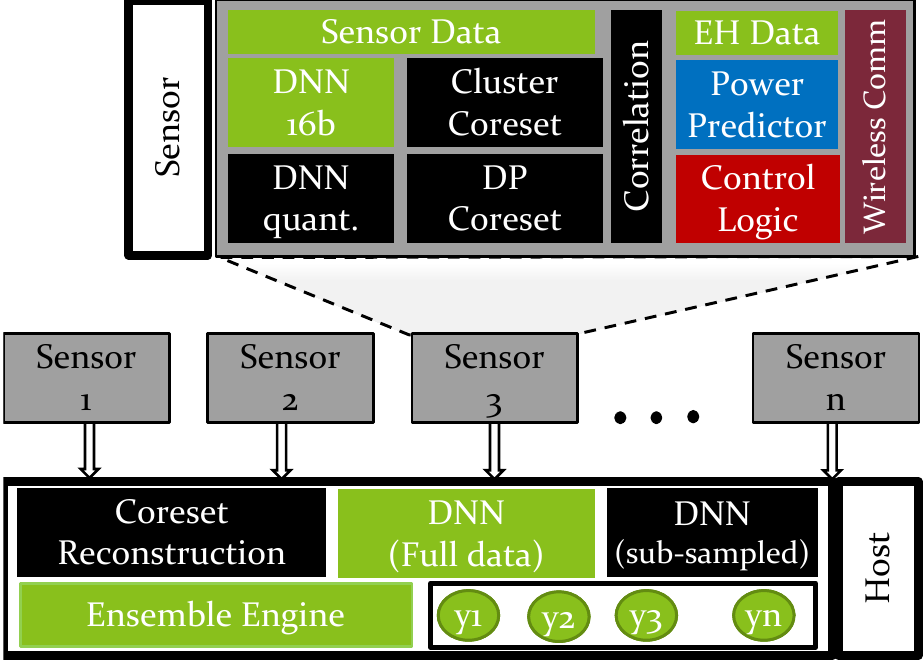}
  \caption{Overall system design of Seeker}
  \label{Fig:system_design}
 \end{wrapfigure}

By leveraging the coreset construction techniques discussed in Section-III, we design \emph{Seeker}. Fig.~\ref{Fig:system_design} gives an overall pictorial representation of the entire sensor host ecosystem of \emph{Seeker} and the various components within it. 
 \textit{Seeker} employs a flexible store and execute method using the ReRAM x-bar hardware described in Origin~\cite{Origin} to perform inference at the edge. \textit{Seeker} extends the Origin approach to provide flexible quantization options to further increase the number of completed inferences. We perform empirical quantitative analysis to determine beneficial quantization vs accuracy tradeoffs (see Fig.~\ref{Fig:quantized-accuracy} : we choose the 16bit and 12bit precision to maximize the number of completed inference requests while minimizing the energy consumption). Moreover, we also augment the sensor with a memoization option so that it does not have to repeat inferences if it encounters similar data, thereby saving substantial energy as well as delivering results with extremely low latency. However, even with all these optimizations, due to the fickle nature of EH, \textit{Seeker} cannot finish all the inferences at the edge and must communicate with a host device. To minimize the data communication overhead between the sensor-node and the host device, \textit{Seeker} utilizes coresets to build representative, yet compressed, forms of the data. 

To cater towards the fickle EH budget, we use two different coreset construction techniques, as described in Section-III: a cheaper, less accurate formation (DP-means) and a recoverable, more expensive formation (K-means). Transmitting coresets rather than raw data greatly improves the energy efficiency of communication to the host, when required, and effectively increases the number of completed inferences, thereby increasing accuracy. Depending on the incoming data and the EH budget, the sensor decides whether to skip compute, perform an inference at the edge, or form a coreset to offload the inference to the host. The host, after obtaining information from multiple sensors, performs any further required computation and uses ensemble learning~\cite{Origin} to give an accurate classification result. Note that, unlike prior EH systems for HAR~\cite{Origin}, the role of the host device is not limited to just result aggregation (or ensembling); rather, the host participates and performs inference when the sensors do not have enough energy to perform it on-sensor and choose to communicate the data (in the form of coresets) to the host. In this section, we will explain in detail the overall execution workflow of the \emph{Seeker} system, followed by the the detailed design of the hardware support to maximize its energy efficiency.

\subsection{Decision Flow: From Sensors to the Host}

\textit{Seeker}'s sensors are equipped with the components described in Fig.~\ref{Fig:system_design}. Each sensor has a data buffer that collects the data points for classification (Implemented using a 60 $\times 3$ FIFO structure of 4Byte cells to store the floating point data. The $\times 3$ indicates the channels for the acceleration component of the three Cartesian axes - X, Y and Z. The moving window is designed using a counter to shift the streaming data.) The sensor also stores one ground truth trace for each activity. The sensor computes the correlation (\circled{1a}) between the stored ground truth and the current data. If the correlation coefficient is $\geq 0.95$ (\circled{1b}) the sensor skips all computation and sends the result directly to the host. Otherwise, the sensor prioritizes local computation and, with the help of a moving average power predictor~\cite{Origin}, predicts whether it can finish the quantized (16 or 12 bit) DNN inference with the combination of stored energy and expected harvesting income (\circled{2a} and \circled{2b}).  If energy is insufficient for DNN inference, the sensor will use coreset formation to communicate the important features to the host, which completes the inference. 

Since the clustering based coresets are typically more accurate then those formed by DP-means, the former are preferred, where possible. We increase the frequency of cluster-based formation by using custom hardware to make clustering more efficient. With the the help of activity aware and recoverable coreset construction and low-power hardware design, we can efficiently communicate inferences or compressed data to the host device with minimum power and latency overheads.

To summarize, \emph{Seeker}, accounting for the available energy budget, considers the following decisions:
\textbf{D0:} Test for data similarity using correlation, and if similarity is found then communicate the results to the host;
\textbf{D1:} DNN at sensor with raw data + Communicate the results to the host (like~\cite{Origin});
\textbf{D2:} Try Quantized DNN inference and communicate the results to the host;
\textbf{D3:} Recoverable coreset construction at the sensor, and communicate the coreset to the host; host runs DNN inference on the reconstructed data; and
\textbf{D4:} Coreset construction at the sensor and communicate the coresets to the host; host runs inference with the compressed coreset data.
\emph{Seeker} takes a sensor-first approach, i.e. to maximize the computations at the senors itself, and then takes an accuracy-optimizing approach, i.e. to maximize the accuracy given the energy budget. Therefore, these decisions are made in the same priority order as they are mentioned ($D0 > D1 > D2 > D3 > D4$). Table~\ref{tab:energy-numbers-table} lists the energy requirements.

\begin{table}[h]
\centering
\resizebox{0.45\textwidth}{!}{%
\begin{tabular}{|l|l|l|l|}
\hline
\textbf{Strategy} & \textbf{Sensor Energy} & \textbf{Communication Energy} & \textbf{Total Energy} \\ \hline
D0 & 0.54  & 8.27  & 8.81  \\ \hline
D1 & 29.23 & 8.27  & 37.5  \\ \hline
D2 & 16.58 & 8.27  & 24.85 \\ \hline
D3 & 0.87  & 15.97 & 16.84 \\ \hline
D4 & 1.07  & 15.97 & 17.04 \\ \hline
\end{tabular}%
}
\caption{Energy breakdown of different Seeker strategies}
\label{tab:energy-numbers-table}\vspace{-16pt}
\end{table}

\subsection{Hardware Support for Energy Efficiency} Energy harvesting brings challenges in both average power levels and power variability. Performing DNN inference under such conditions often limits exploitation of inherent DNN parallelism within the energy budget. Therefore, many prior works~\cite{ResiRCA, Origin} use a ReRam x-bar to perform on-sensor-node DNN inference. \textit{Seeker}'s inference engine follows the design proposed in ResiRCA~\cite{ResiRCA} and modifies it to cater towards new quantization requirements: We have two different instances of the ReRAM x-bar in our system - one for the 16bit model and one for the 12-bit model. The nonvolatile nature of the RERAMs helps in performing intermittent computing with the harvested energy. Moreover, techniques like loop tiling and partial sums~\cite{ResiRCA} can further break down the computation to maximize forward progress with minimum granularity.  

While DNN inference is already accelerated in the sensor node, the addition of a general purpose processor for coreset computation would be energy inefficient. Specifically, the computation requirement is fixed and does not require the overheads to support generality. Hence, we add low power, low latency coreset construction hardware. Both the coreset construction algorithms follow a sequence of multiply and add/subtract operations followed by averaging, and hence can be simply designed with few logic units. Moreover, as we are operating with lower data volume, these operations can be parallel (for example, the clustering hardware simultaneously works on all the cluster formations). The bigger challenge is posed by the requirement of a variable number of iterations for these algorithms to converge, the number of clusters/samples required, etc. 
To efficiently design the hardware and configure its parameters, we run several experiments and empirically arrive at the following conclusions: \textbf{1.} The clustering finishes within 4 iterations of the k-means algorithm and for DP-means it takes up to 7 iterations. \textbf{2.} None of the clusters have more than 16 points during any clustering. \textbf{3.} We need not store all the points in either cases at every iteration, rather clustering hardware need to store the sum, the radius of the cluster, the number of points per cluster and, DP-means hardware needs just the point. \textbf{4.} Storing the radius helps in easily selecting the points in the subsequent iterations.

%% file: src/05Evaluation.tex
\begin{figure*}[htbp]
    \subfloat[Accuracy with MHEALTH dataset]
    {
    \includegraphics[width=0.48\linewidth, height=3.5cm]{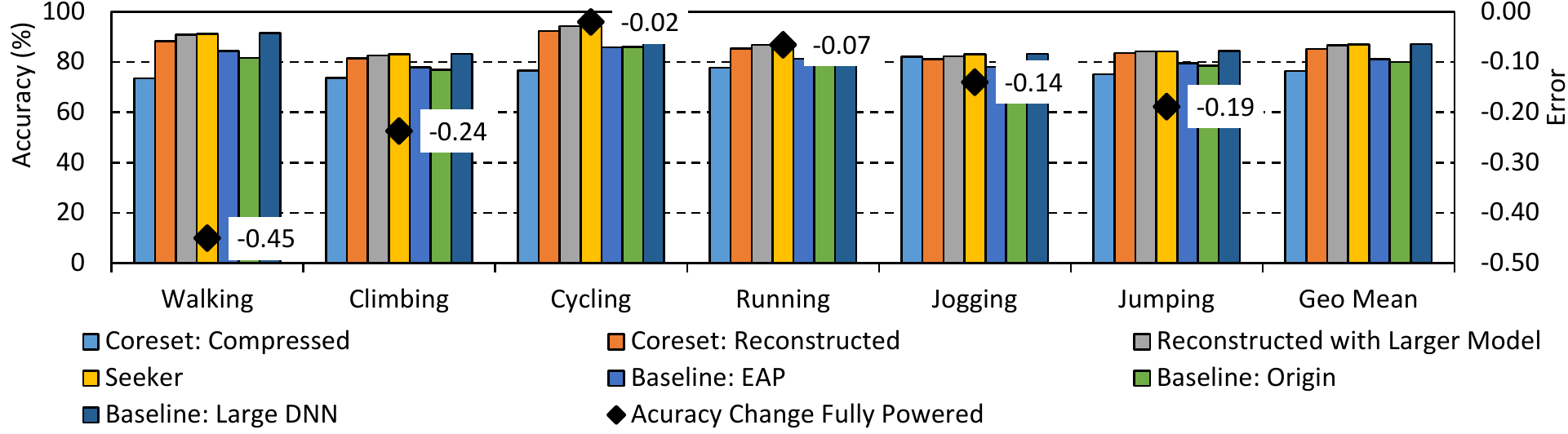}\label{Fig:MHEALTH-accuracy}
    }
    \hfill
    \subfloat[Accuracy with PAMAP dataset]
    {
    \includegraphics[width=0.48\linewidth, height=3.5cm]{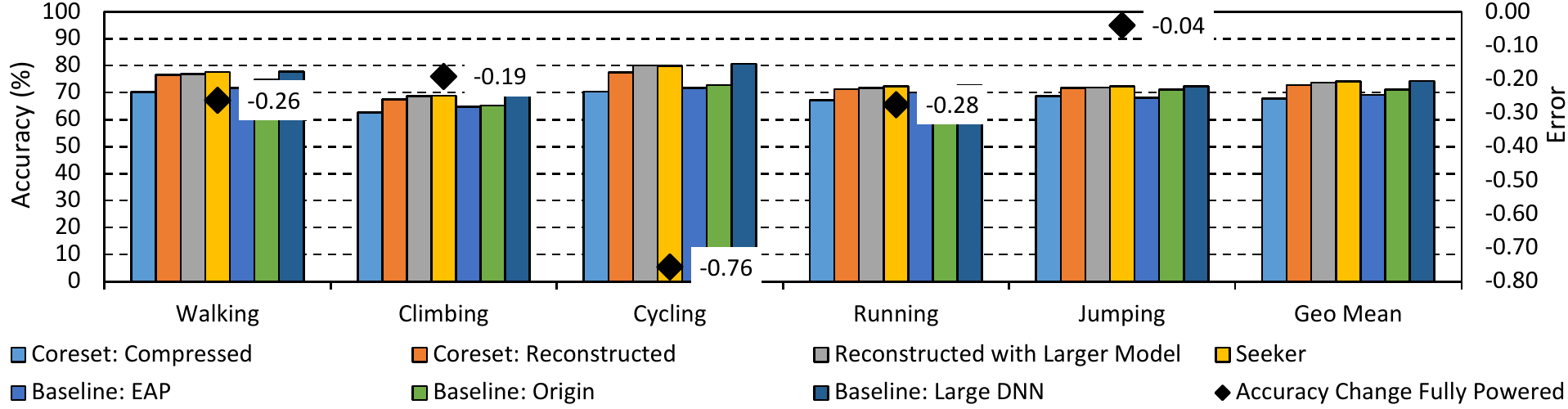}\label{Fig:PAMAP-accuracy}
    }
    
    \subfloat[Data volume with activity aware clustering]
    {%
     \includegraphics[clip,width=0.3\linewidth, height=3cm]{figs/12.pdf}%
    \label{Fig:clusters-datavol}
    }
    \hfill
    \subfloat[Fraction of inferences completed with different EH sources]
    {%
     \includegraphics[clip,width=0.3\linewidth, height=3cm]{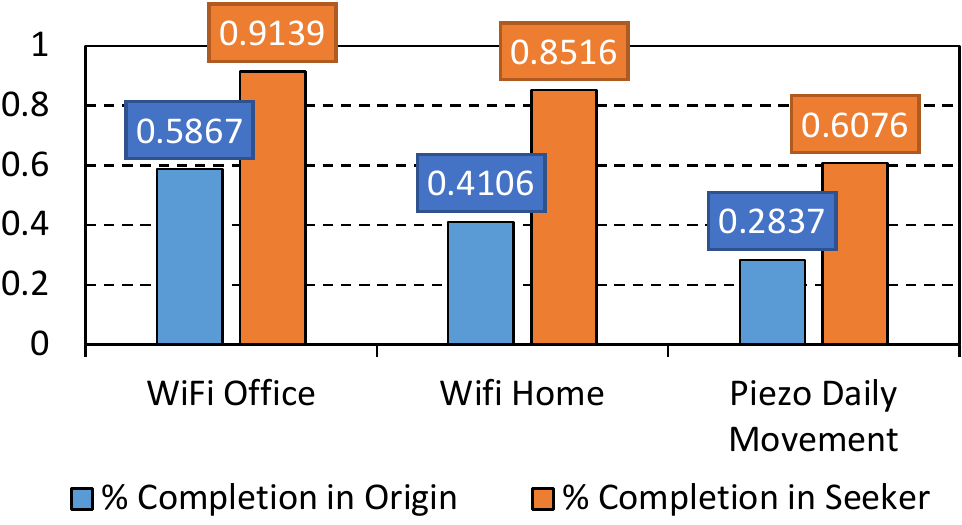}%
    \label{Fig:Sensitivity-fracCompleted}
    }
    \hfill
    \subfloat[Distribution of compute off-load to different components]
    {%
  \includegraphics[clip,width=0.3\linewidth, height=3cm]{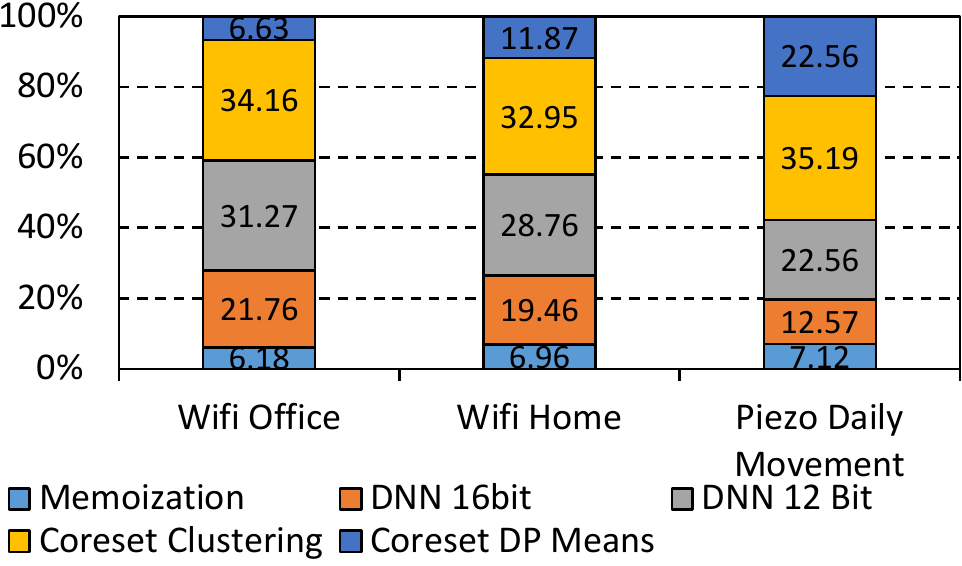}%
  \label{Fig:Sensitivity-fracOffloaded}
    }
    \caption{Accuracy and communication efficiency of \emph{Seeker} with different data sets and its sensitivity towards various EH sources.}
    \vspace{-0.1cm}
    \label{fig:AllAccuracy}  
    \vspace{-0.5cm}
\end{figure*}

In this section, we describe our methodology for evaluating \textit{Seeker}, compare it against three accuracy baselines with different energy-optimization assumptions, and perform a sensitivity study across various harvested energy sources.

\subsection{Evaluation Methodology} Our evaluation design for human activity recognition uses three different sensors located at left ankle, right arm, and chest. Each sensor consists of the following components: sensing element (an Inertial Measurement Unit, which collects acceleration data), an energy harvester (we use RF energy collected from a WiFi source), three DNN inference engines (a 16bit and a 12bit working on raw sensor data, and a 16bit working on the compressed coreset data), a correlation computation engine, two coreset computation engines (for DP-means, and for cluster-based coreset), power predictor, and low energy communication option based on IEEE 802.15.6, which is specifically designed for wireless body area networks. 
The specifics of the energy-harvesting mechanism producing the power trace are beyond the scope of this work. We use two different datasets, MHEALTH~\cite{mHealth,mHealthDroid}, and
PAMAP2~\cite{PAMAP21,PAMAP22}, for our evaluation. The DNNs are  trained on the training data-sets using Keras~\cite{keras}. 

We choose three points of comparison for our accuracy evaluation. \textbf{Baseline-1 (Baseline: Large DNN)} consists of a full precision (without any pruning) DNN built on the lines of~\cite{HARHaChoi}. \textbf{Baseline-2 (Baseline: EAP)} uses state of the art optimizations~\cite{netadapt} on Baseline-1 to design a power-aware DNN tuned to for the average harvested power of our EH source. \textbf{Baseline-3 (Baseline: Origin)} uses the system design proposed in~\cite{Origin}. Baseline-1 and Baseline-2 run on fully powered systems where as Baseline-3 runs on the same EH source as \textit{Seeker}. For communication, we consider a system which transmits the entire raw data to the host as the baseline.   

\subsection{Analysis of Results}


Fig.~\ref{Fig:MHEALTH-accuracy} and Fig.~\ref{Fig:PAMAP-accuracy} show the accuracy of various policies described in Section~\ref{sec:4}, along with the accuracy of \emph{Seeker} - which applies all policies together, along with ensemble learning.
Fig.~\ref{Fig:clusters-datavol} shows the normalized data communication volume with different numbers of clusters, along with activity-aware clustering. We make the following observations:
\squishlist
\item With RF as the EH source, \textit{Seeker} finishes or offloads about 91.4\% of the inferences scheduled on the sensors, compared to 58.7\% in recent work~\cite{Origin}. The percentage varies with different EH sources, but the relative trend persists. 
\item We observe that the compressed clusters give less accuracy, evidently because of the loss of features during the coreset formation. However, with the reconstructed clusters, the accuracy reaches 86.8\% compared to 76.4\% for the former.
\item \emph{Seeker}, thanks to synergistic computation, achieves 87.05\% accuracy with MHEALTH (0.18\% less accurate than a DNN running at full precision with full power) data set and similarly reaches 74.2\% accuracy on the PAMAP2 data set. This gives us $\approx7\%$ more accuracy than~\cite{Origin} on MHEALTH and $\approx3\%$ more accuracy for PAMAP2 dataset.   
\item With activity-aware coreset construction, \emph{Seeker} reduces the communication volume by $8.9\times$ compared to the baseline. 
\item The proposed coreset construction hardware design is simulated using Design Compiler and takes 0.1$mm^2$ area and consumes 130$\mu W$ of power. 
\squishend

\subsection{Sensitivity Study} Our experiments with other EH sources show the versatility of \emph{Seeker}, which outperforms a HAR classifier designed for EH~\cite{Origin} across multiple (piezo-electric, RF) harvested energy sources and demonstrate that it can easily be scaled to work with any number of sensors. Fig.~\ref{Fig:Sensitivity-fracCompleted} shows the comparison of scheduled inferences completed while using \emph{Seeker} and the state-of-the-art approach~\cite{Origin}. Further, we demonstrate how \emph{Seeker} leverages \textbf{all} the proposed design components (including memoization, DNN inference engine and coreset engine) to complete maximum compute at the edge and offload minimum to the host device. Fig.~\ref{Fig:Sensitivity-fracOffloaded} shows the compute breakdown among components under different EH sources.


%% file: src/06Conclusion.tex
As systems utilizing energy harvesting edge devices are tasked with increasing complex tasks, like HAR, both system and node designs must respond with targeted, efficiency-maximizing optimizations. Our proposal, \emph{Seeker}, synergizes wearable sensor nodes with personal mobile devices by intelligently distributing computations among them while minimizing the communication overhead. Our experiments show that, by leveraging coreset techniques in data reduction and tuning these techniques for application-aware properties, \emph{Seeker} can provide better accuracy ($86.8\%$), even while limited to harvested power, than state of the art energy-optimized DNNs running on a fully powered device ($81.2\%$). Furthermore, it also outperforms a recent HAR system designed specifically for EH-WSNs. Collectively, the optimizations in \emph{Seeker} can reduce communication traffic by $8.9\times$ while improving inference accuracy, demonstrating the potential of holistic system/node/application optimization in the HAR space.